\documentclass[%
 reprint,
superscriptaddress,
 amsmath,amssymb,
 aps,
]{revtex4-1}

\usepackage{graphicx}
\usepackage{dcolumn}
\usepackage{bm}
\usepackage{hyperref}

\usepackage{caption} 

\begin{document}

\preprint{APS/123-QED}

 \title{
Testing scalar-tensor-vector gravity theory via black hole photon rings
}%




\author{Qiao Yue}
\author{Zhaoyi Xu}

 \author{Meirong Tang}
 \email{Electronic address: tangmr@gzu.edu.cn(Corresponding author)
}
\affiliation{%
College of Physics, Guizhou University,\\
 Guiyang 550025, China
}%
\date{\today}

\begin{abstract}
In the scalar-tensor-vector gravitational framework, this paper systematically studies the photon ring and shadow structure of the Reissner-Nordström black hole. In this model, the black hole is determined by the \( MOG\) parameter $\alpha $ and the charge $Q$. Through calculations, it is found that as $\alpha $ increases, the event horizon radius $r_h$, the photon sphere radius $r_{ph}$, and the critical impact parameter $b_{ph}$ all gradually increase. On the other hand, as $Q$ increases, all three decrease. The innermost stable circular orbit radius $r_{isco}$ also exhibits the same monotonic behavior. Backward ray-tracing shows that as \( Q \) increases, the impact parameter \( b \) interval between the lensing ring and the photon ring widens; \( b_{\text{ph}} \) is non-degenerate, and the photon ring radius is uniquely determined by the parameters $\alpha $ and \( Q \). Using the combined $ EHT $ constraints on $ SgrA^* $ and $ M87^* $, we obtain upper and lower bounds on $\alpha $ and \( Q \). For $Q = 0$, $0.5$, and $1$, the allowed ranges are  $\alpha \in \left [0 , 0.06 \right ],\left [ 0,0.11 \right ]$ , and $\left [ 0.19, 0.36 \right ]$, respectively. Radiative simulations for three toy models show that, at fixed \( Q \), larger $\alpha $ yields a larger, non-degenerate photon ring; Only when the values of $\alpha $ and \( Q \) are small can it approach the Schwarzschild case, providing a key computational basis for testing modified black holes. This result offers a non-degenerate observational criterion for distinguishing quantum gravity models, which is consistent with current $ EHT $ data. Future observations with $ ngEHT $ and multi-band polarization can further test this. This feature strongly suggests that investigating the fine structure of the photon ring of a Reissner–Nordström black hole in scalar–tensor–vector gravity offers a distinctive optical diagnostic, enabling potential tests of quantum-gravity signatures and providing insight into the black hole’s intrinsic properties.
\begin{description}
\item[Keywords]
 Modified black holes, Scalar-tensor-vector gravity, Photon ring, Black-hole shadow, Event horizon, Quantum-gravity effects 
\end{description}
\end{abstract}

\maketitle

\section{Introduction}\label{1.0}

Black holes are not only a hallmark prediction of general relativity ($GR$), but also a unique cosmic laboratory that connects the microscopic quantum world with the macroscopic structure of the universe, providing an opportunity to explore the behavior of physical laws under extreme conditions \cite[e.g.][]{MisnerThorneWheeler1973,CouperHenbest1996,LIGOScientific:2016aoc}, They occupy a central place in theoretical physics and cosmology. In $ 2019 $, the Event Horizon Telescope ($ EHT $) collaboration produced the first direct image of the supermassive black hole at the center of the giant elliptical galaxy $ M87^* $ and released the first-ever photograph of a black hole \cite{EventHorizonTelescope:2019dse}. This milestone not only delivered persuasive observational support that supermassive black holes inhabit galactic nuclei and power active galactic nuclei, but also opened a new avenue for testing gravity in the extreme strong-field regime and on an unprecedented mass scale \cite[e.g.][]{EventHorizonTelescope:2019uob,EventHorizonTelescope:2019jan,EventHorizonTelescope:2019ths,EventHorizonTelescope:2019pgp,EventHorizonTelescope:2019ggy}. In $ 2022 $, the $ EHT $ collaboration published the first event-horizon-scale image of the supermassive black hole $ SgrA^* $ at the Galactic Center, based on observations conducted with the global very-long-baseline interferometry ($VLBI$) array \cite{EventHorizonTelescope:2022wok}. This result provides direct observational evidence that the “compact mass concentration” at the Galactic center is indeed a black hole, furnishes a benchmark sample for studying the co-evolution of super-massive black holes with their host galaxies (the Milky Way), stellar-dynamical feedback, and black-hole accretion physics, and fills the empirical void in $ SgrA^* $ research \cite[e.g.][]{EventHorizonTelescope:2022wok,Lu:2018uiv,Grigorian:2024rsn,EventHorizonTelescope:2022wkp,EventHorizonTelescope:2022apq,EventHorizonTelescope:2022exc,EventHorizonTelescope:2022urf,EventHorizonTelescope:2022xqj}.

In general relativity, the photon ring is a nested structure around a black hole composed of photons in critically bound orbits \cite{Gralla:2019xty}, lying within the photon sphere (a spherical surface with a radius approximately $ 1.5 $ times the Schwarzschild radius, where $ r= \frac{3GM}{C^{2} } $) \cite[e.g.][]{Gogoi:2024vcx,Johnson:2019ljv,Himwich:2020msm}]. In this region, the strong spacetime curvature traps photons incident at specific angles into closed orbits. Such photons may execute several revolutions around the black hole prior to escaping or being absorbed by the event horizon. Their geodesics give rise to a hierarchy of concentric rings, yielding an “infinite-mirror”-type structure\cite[e.g.][]{Wang:2023vcv,Wang:2022yvi,Chael:2021rjo}. Photon rings are of great significance in the measurement of black hole spin, tests of strong-field gravity, black hole accretion physics, and exploration of quantum gravity \cite[e.g.][]{Beckwith:2004ae,Gao:2023mjb,Luminet:1979nyg,Gan:2021xdl,Takahashi:2004xh,Yang:2024utv,Zeng:2025kqw,Wang:2025ihg,Zeng:2024ptv,Zeng:2023zlf,Zeng:2022pvb,Zeng:2022fdm,Yue:2025fly}, and constitute a core probe in modern black hole astrophysics. Observations of photon rings require extremely high angular resolution at the microarcsecond level, which was first achieved for direct imaging by the $ EHT $ through global millimetre-wave very long baseline interferometry \cite[e.g.][]{EventHorizonTelescope:2019dse,EventHorizonTelescope:2022wok,EventHorizonTelescope:2021bee}. Theoretical modeling combines black hole spacetime metrics, radiation transfer, and magnetohydrodynamic simulations. It calculates the propagation paths of photons through general relativistic ray-tracing algorithms, predicts the observational characteristics of photon rings, and provides a quantitative tool for constraining black hole parameters \cite[e.g.][]{Gralla:2019xty,Johnson:2019ljv,Dexter:2016cdk,Moscibrodzka:2017lcu,Chael:2024gvx,Gammie:2003rj,McKinney:2004ka,Dhruv:2024igk,Jimenez-Rosales:2021ytz}.

To explain galactic rotation curves, galaxy cluster dynamics, and cosmological phenomena while avoiding the introduction of non-baryonic dark matter, the academic community previously proposed the Non-symmetric Gravitational Theory ($NGT$) \cite{Moffat:1994hv} and the Metric-Scalar-Tensor Gravitational theory ($MSTG$) \cite{Moffat:2004bm}. To build a more complete relativistic gravitational framework, Moffat further proposed the Scalar-Tensor-Vector Gravity theory ($STVG$) \cite{Moffat:2005si}, hereafter denoted as Modified Gravity ($MOG$). By allowing a dynamical scalar field to evolve through spacetime-augmented by strong infrared-renormalization effects-the theory replaces dark matter in explaining cosmic structure, while a neutral vector field, whose special properties (no radiative coupling, zero pressure) prevent it from damping fluctuations in the cosmic microwave background ($CMB$). At present, this theory has been applied in the fields of galaxy rotation curve fitting \cite[e.g.][]{Brownstein:2005zz,Moffat:2004bm}, galaxy cluster observation and gravitational lensing \cite[e.g.][]{Brownstein:2005dr,Saydullayev:2025oop,Tizfahm:2024rcj,Moffat:2008gi}, solar system and compact object observation tests \cite[e.g.][]{Bertotti:2003rm,Adelberger:2003zx}, as well as quantum gravity and effective field theory \cite[e.g.][]{Reuter:2004nx,Prokopec:2005fb}. This theory has yielded key contributions to the connection with $GR$ through multi-faceted studies. It points out that observational data in strong gravitational fields represent the core for distinguishing $STVG$ from other gravitational theories, and proposes subsequent research directions such as refining parameter evolution and fitting $CMB$ perturbations, thereby providing theoretical guidance for relevant experimental tests.

This paper conducts a study based on the Reissner-Nordström ($RN$) black hole model within the framework of $STVG$. By deriving the geodesic equations, constructing the effective potential functions, and combining the backward ray-tracing technique, we perform numerical simulations of the propagation behavior of photons in strong gravitational fields. Furthermore, we investigate the distinctive optical signatures of these black holes in order to probe their spacetime geometry and dynamical behavior. Our analysis indicates that, within the $STVG$ framework, the photon-ring fine structure of the $RN$ black hole can act as a characteristic optical diagnostic, potentially sensitive to quantum-gravity imprints and informative of intrinsic black-hole properties. This, in turn, provides a quantitative avenue for confronting black-hole physics and candidate quantum-gravity scenarios with observations.

The remainder of this paper is organized as follows. Section \ref{2.0} presents the $RN$ black hole in the $STVG$ framework and discusses how the $MOG$ parameter $\alpha$ and the electric charge \( Q \) modify the black-hole spacetime geometry. Section \ref{3.0} derives the photon null-geodesic equations and the corresponding effective potential, clarifies how $\alpha$ and \( Q \) regulate photon orbital dynamics, and identifies the principal orbital quantities, including the photon-sphere parameters. Section \ref{4.0} analyzes the fine structural characteristics of the photon ring: first, the bending characteristics of photon trajectories and their interaction modes with the accretion disk under different values of $\alpha $ and \( Q \) are investigated by solving the null geodesic equations. Second, the $ EHT $ observational data are employed to constrain the parameters $\alpha $ and \( Q \); finally, three emission models are established to simulate the observational characteristics of the photon ring, and the influence laws of $\alpha $ and \( Q \) on the radius, brightness distribution and imaging morphology of the photon ring are quantitatively analyzed. Section \ref{5.0} summarizes the unique properties of the photon ring structure and discusses the scientific significance of its role as a probe for quantum gravitational effects. Geometric units with \( c = G = 1 \) are adopted for all calculations in this paper, where the mass parameter is set to \( M = 1 \) and the metric signature is \( (-, +, +, +) \).

\section{Reissner-Nordström modified gravity black hole}\label{2.0}
In this section, we examine the $RN$ black hole solution in modified gravity by focusing on its fundamental properties. The model was introduced by Isomiddin-Nishonov et al. in the presence of the vector field $\phi_{\mu}$, with the aim of exploring the interaction between test particles and the $STVG$ field \cite{Nishonov:2025hxz}. For the exterior region, we consider a static and spherically symmetric spacetime written in spherical coordinates, described by the line element\cite{Nishonov:2025hxz}: 
\begin{equation}\label{1}
ds^{2} = -f(r) dt^{2} + \frac{1}{f(r)} dr^{2} + r^{2} ( d\theta ^{2} + sin^{2}\theta d\varphi ^{2} ),
\end{equation}
where
\begin{equation}\label{2}
f(r)=1-\frac{2(1+\alpha )M}{r} +(1+\alpha ) \frac{(\alpha M^{2}+Q^{2})}{r^{2} } .
\end{equation}
Here, $\alpha $ is the $ MOG $ parameter \cite{Wondrak:2018fza} and \( Q \) is the electric charge, which characterize the deviation of the $ STVG $ theory from the Schwarzschild solution.

It can be seen from Eqs.\eqref{1} and \eqref{2} that the solution reduces to the Schwarzschild black hole solution when both parameters $\alpha $ and \( Q \) are set to zero. 

The event horizon of this black hole satisfies the relation \cite{Nishonov:2025hxz}:
\begin{equation}\label{3}
r_{\pm } = (1+\alpha ) M \pm \sqrt{(\alpha +1) M^{2}- Q^{2}} .
\end{equation}
For an extremal charged black hole, the critical charge (at which the event horizon meets the extremal condition) satisfies \cite{Nishonov:2025hxz}
\begin{equation}\label{4}
\left | Q \right |_{extr } = M ,\left (r _{h}  \right )_{min} = (1+\alpha ) M  .
\end{equation}

\begin{figure}[]
\includegraphics[width=0.5\textwidth]{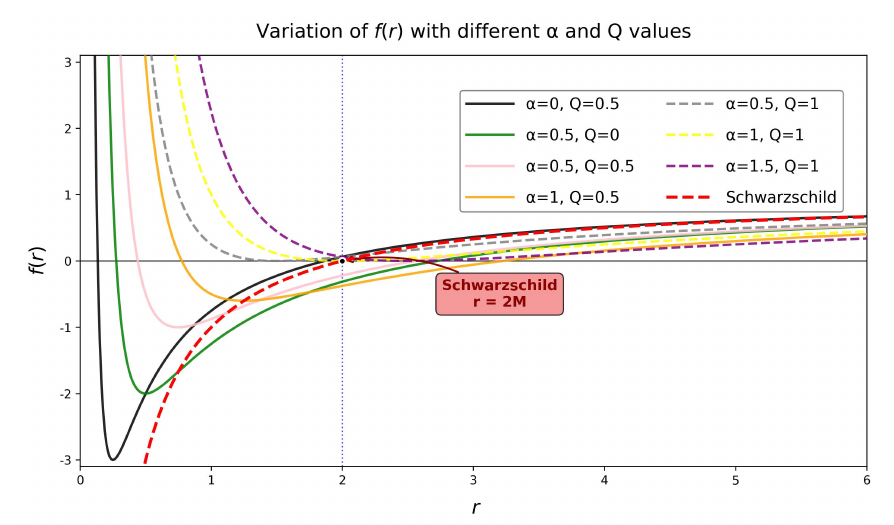}
\caption{
Metric function \(f(r)\) for various $\alpha$ and \( Q \) at \( M = 1 \), with the Schwarzschild case shown for comparison.}
\label{a}
\end{figure}

Therefore, the possible event horizons in this metric can be obtained by finding the solutions \(r=r(\alpha , Q)\) to the equation
\begin{equation}\label{5}
f(r)=0.
\end{equation}
The dependence of the radial metric function \(f(r)\) on the $MOG$ parameter $\alpha$ and the electric charge \( Q \) is presented in Fig.\ref{a} for multiple parameter settings.

It follows from Fig.\ref{a} that the RN solution in modified gravity exhibits parameter-dependent horizon properties. If $\alpha$ and \( Q \) are large enough, the lapse function \(f(r)\) does not cross zero and remains positive, signaling that the event horizon disappears. Accordingly, the presence of an event horizon constrains the parameters to $\alpha \in [0,0.5]$ and $Q\in [0,1]$. Moreover, at radii well outside the horizon, the spacetime differs appreciably from Schwarzschild, implying that weak-field tests provide only limited constraints on these modifications. A meaningful assessment of the effects of $\alpha$ and \( Q \) thus requires entering the strong-field gravitational regime \cite{Moffat:2005si}.

\begin{figure}[]
\includegraphics[width=0.5\textwidth]{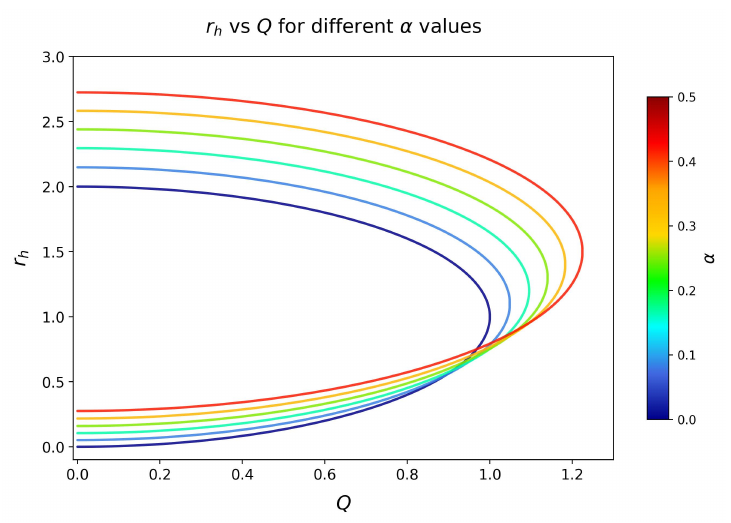}
\caption{
Variation of \(r_{h}\) with the $MOG$ parameter $\alpha$ and the electric charge \( Q \).}
\label{b}
\end{figure}

As presented in Fig.\ref{b}, the dependence of the horizon radius \(r_{h}\) on the parameters $\alpha$ and \( Q \) is examined for several representative parameter pairs.

It is evident from Fig.\ref{b} that the horizon radius \(r_{h}\) responds sensitively to the $MOG$ parameter $\alpha$, reflecting the manner in which modified gravity adjusts the horizon location. Moreover, varying \( Q \) leads to distinct horizon radii, and the family of curves corresponding to different charges provides a direct visualization of the charge-induced changes in the horizon scale.

\section{Geodesic Equations and Effective Potentials of Reissner-Nordström Black Holes in Modified Gravity}\label{3.0}
Deriving the geodesics and effective potentials is crucial for elucidating both the geometric features of black-hole spacetimes and the dynamics of particles propagating within them. Within the framework of general relativity, the null geodesic equations for photons are derived from the metric \eqref{1} by applying the variational principle in light of the motion characteristics of particles in curved spacetime. Due to spherical symmetry and stationarity, the spacetime possesses two Killing vectors, leading to two conserved quantities: the energy $ E $ and the angular momentum $ L $. Restricting to the equatorial plane $\theta =\frac{\pi }{2} $, the null-geodesic equations take the form (see Appendix for details):
\begin{equation}\label{6}
(\frac{du}{d\varphi } )^{2} = \frac{1}{b^{2} } - u^{2} f(\frac{1}{u} ).
\end{equation}
where $ u=\frac{1}{r} $  and $ b=\frac{L}{E} $ denote the impact parameter. The effective potential is defined as:
\begin{equation}\label{7}
V_{eff}(r) = u^{2} f(\frac{1}{u} ).
\end{equation}
Herein, we refer to the circular null geodesics as photon spheres, which describe the trajectories of photons; The local extrema of the effective potential occur where the following condition holds \cite{Gan:2021pwu}:
\begin{equation}\label{8}
V_{eff}( r_{ph} ) = \frac{1}{b_{ph}^{2} } .
\end{equation}
and
\begin{equation}\label{9}
V_{eff}^{'}( r_{ph} ) = 0.
\end{equation}

In Eqs. above, the photon sphere is characterized by its radius \( r_{\text{ph}} \) and the critical impact parameter \( b_{\text{ph}} \). The Schwarzschild relation between these quantities is illustrated in Fig.\ref{3}. In Fig.\ref{4}, we plot the radial dependence of the effective potential \( V_{\text{eff}} (r) \) for several representative pairs of $\alpha$ and \( Q \).
\begin{figure}[]
\includegraphics[width=0.5\textwidth]{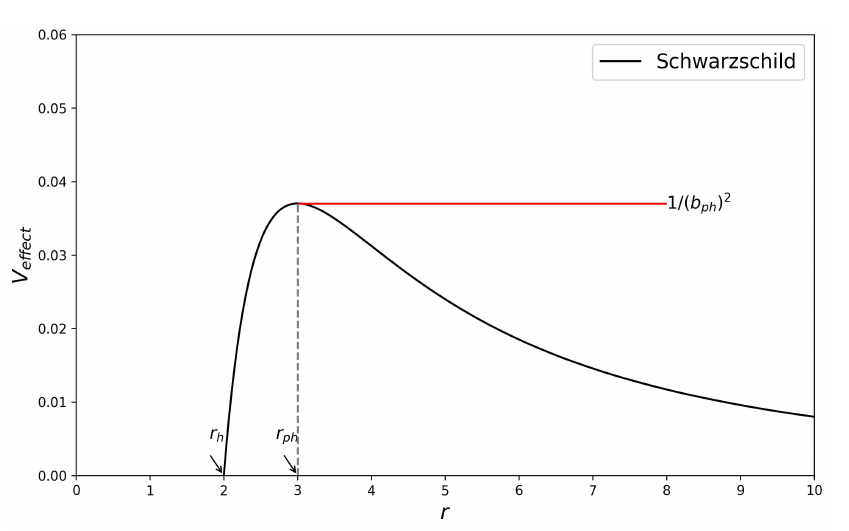}
\caption{
Effective potential \( V_{\text{eff}} (r) \) for a Schwarzschild black hole and its correspondence to the photon-sphere radius \( r_{\text{ph}} \) and the critical impact parameter \( b_{\text{ph}} \)}.
\label{c}
\end{figure}
\begin{figure*}[]
\includegraphics[width=1\textwidth]{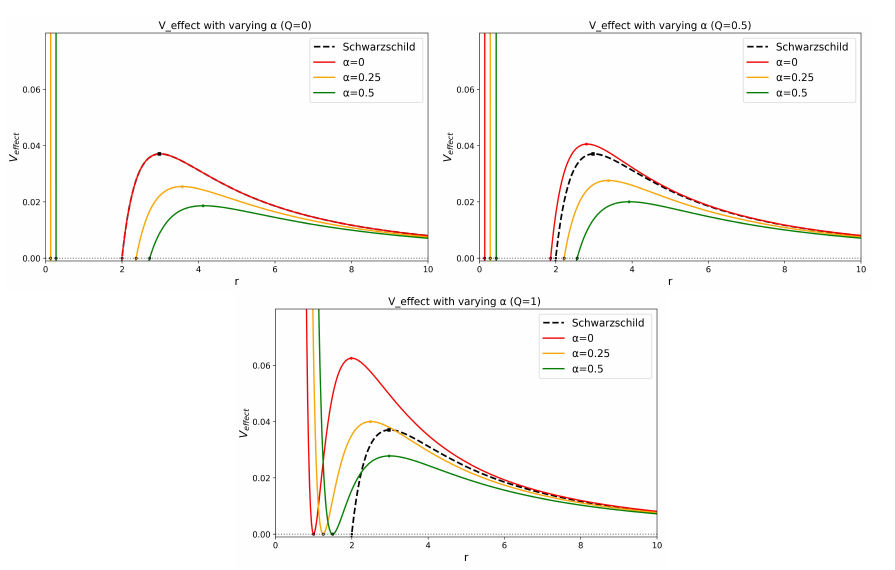}
\caption{
$V_{eff} (r)$ - $r$ curves for $ Q=0$ (upper-left), $Q=0.5$ (upper-right), and $Q=1$ (lower-center), shown for $\alpha=0$, $0.25$, and $1$, with the Schwarzschild case added as a benchmark.}
\label{d}
\end{figure*}

For a Schwarzschild black hole, Fig.\ref{c} shows \( r_h = 2 \) for the event horizon and \( r_{\text{ph}} = 3 \) for the photon sphere.

It can be seen from Fig.\ref{d} that in the region far from the central mass, the effective potential behaves consistently with the Schwarzschild solution regardless of the values of $\alpha $ and \( Q \). However, in the short-distance regime close to the central mass, the effective potential deviates from the Schwarzschild solution; furthermore, the larger the value of \( Q \), the more the overall peak shifts to the left (i.e., both the event horizon radius \( r_h \) and the photon sphere radius \( r_{\text{ph}} \) decrease gradually).

The maximum of the effective potential corresponds to the radius of the unstable photon sphere, where a particle will deviate from its trajectory upon a slight perturbation; in contrast, the minimum corresponds to the radius of the stable photon sphere, where a particle tends to return to its original position after being perturbed. In the research of astrophysics and gravitational theory, unstable photon rings have emerged as a key focus of photon ring studies due to their ubiquitous existence in strong gravitational fields, high sensitivity to spacetime structures, with implications for astronomical observations, notably accretion-disk radiation and gravitational-lensing effects \cite[e.g.][]{Gan:2021pwu,Meng:2024puu,Zhong:2024ysg,Cunha:2017eoe}.

\section{Fine Structure of the Photon Sphere Orbits}\label{4.0}
The fine structure of photon-sphere orbits directly reflects the strong-field gravity of a black hole. Near the event horizon, photons propagate along null geodesics and generate a hierarchy of nested sub-rings, whose radii, brightness, and even stability are controlled by intrinsic black-hole parameters such as mass and spin. In this section, we apply backward ray tracing to study photon propagation around the $RN$ black hole in the $STVG$ framework and its interaction with an equatorial thin accretion disk. The disk is modeled as geometrically thin, and the observer is placed along the direction of the black hole’s north pole. Owing to strong gravitational bending, photon trajectories can intersect the disk plane repeatedly, producing intricate, partially overlapping paths. We classify the rays by the number of geodesic-disk intersections and quantify how each class contributes to the total observed intensity. Based on this classification, we compute the cumulative brightness distribution, reconstruct the black-hole image for a background-free emitting disk, and examine how multiple path windings shape the shadow and photon-sphere substructure.

\subsection{null geodesic}\label{4.1}
The null-geodesic equation \eqref{6} implies a direct correspondence between photon trajectories and the impact parameter $b$. In particular, for geodesics traveling parallel to the north-polar axis, Ref. \cite{Gralla:2019xty} shows that the admissible impact-parameter ranges are classified by the number of disk crossings. Specifically, one has:
\begin{equation}\label{10}
n = \frac{\varphi }{2\pi }.
\end{equation}
where \(\varphi\) denotes the total variation of the polar angle for a null geodesic over its entire trajectory in the polar coordinate system. For \(b < b_{\text{ph}}\), the total azimuthal angle variation of the trajectory outside the event horizon is given in Refs. \cite[e.g.][]{Gralla:2019xty,Peng:2020wun} as follows:
\begin{equation}\label{11}
\varphi = \int_0^{u_h} \frac{1}{\sqrt{\frac{1}{b^2} - f\left(\frac{1}{u}\right) u^2}}du.
\end{equation}
Let $u_{h} \equiv r_{h}^{-1}$ , with \( r_h \) the radius of the outer event horizon. For the regime \( b > b_{\text{ph}} \), the net azimuthal angle change becomes
\begin{equation}\label{12}
\varphi = 2 \int_0^{u_{\text{max}}} \frac{1}{\sqrt{\frac{1}{b^2} - f\left(\frac{1}{u}\right) u^2}} du.
\end{equation}
In this notation, $u_{max} \equiv r_{min}^{-1}$ , where \( r_{\text{min}} \) is the smallest radius reached by the photon trajectory. For the radiation images formed near a black hole, the received intensity is determined by the intersection count between null geodesics and the accretion disk. Therefore, $ N = n(b)$ and it is piecewise specified by the criteria summarized in the references \cite[e.g.][]{Gralla:2019xty,Peng:2020wun}
\begin{equation}\label{13}
n(b) = \frac{2m - 1}{4}, \quad m = 1, 2, 3, \cdots.
\end{equation}

For a fixed value of \( m \), the equation yields two distinct roots, denoted \( b_m^- \) and \( b_m^+ \), which correspond to the smaller and larger solutions, respectively. Using this, the ways in which null geodesics intersect the accretion-disk plane can be grouped into three regimes:

Regime 1: the null geodesic crosses the disk plane only once, representing direct emission (\( n < 3/4 \)).

Regime 2: the null geodesic crosses the disk plane twice, corresponding to lensing-ring emission (\( 3/4 < n < 5/4 \)).

Regime 3: the null geodesic intersects the disk plane three or more times, associated with photon-sphere emission (\( n > 5/4 \)).

\begin{figure}[]
\includegraphics[width=0.5\textwidth]{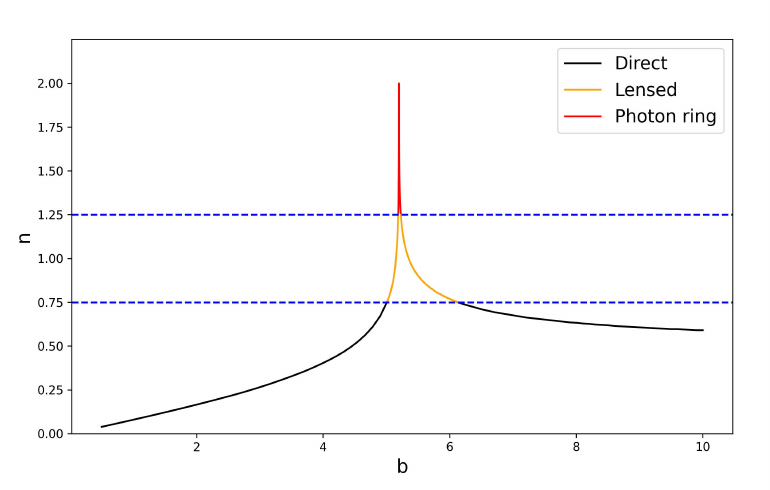}
\caption{
Curve of \( n \) versus the impact parameter \( b \) for a Schwarzschild black hole, where black denotes Category $1$, orange denotes Category $2$, and red denotes Category $3$.}
\label{e}
\end{figure}

The relationship between \( n \) and the impact parameter \( b \) for a Schwarzschild black hole is shown in Fig.\ref{e}.

According to the results shown in Fig.\ref{e}, the classification of these light rays is given as follows:

(a)Direct emission: In the regime \( n < 3/4 \), the impact parameter \( b \) values is \( (0, b_2^-) \bigcup (b_2^+, \infty) \);

(b)Gravitational lens emission: For \( 3/4 < n < 5/4 \), the corresponding range of \( b \) values is \( (b_2^-, b_3^-) \bigcup (b_3^+, b_2^+) \);

(c)Photon ring emission: For \( n > 5/4 \), the corresponding range of \( b \) values is \( (b_3^-, b_3^+) \).

\begin{figure*}[]
\includegraphics[width=1\textwidth]{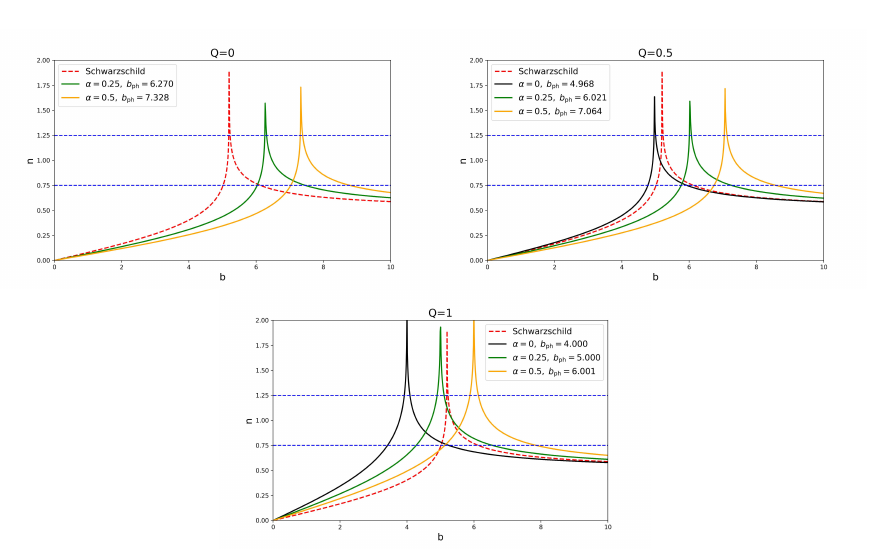}
\caption{
Curves of \( n \) versus the impact parameter \( b \) for $Q=0$ (top left), $Q=0.5$ (top right), $Q=1$ (bottom center), with $\alpha=0$, $0.25$, $0.5$ and the Schwarzschild solution (red) included for comparison.}
\label{f}
\end{figure*}

Meanwhile, by applying the null-geodesic equation in \eqref{6}, the $n-b$ relation can be derived, as specifically illustrated in Fig.\ref{f}. Here, a comparative analysis is conducted for the cases where the value range of \( Q \) is $[0,1]$, the value range of $\alpha $ is $[0,0.5]$, and the Schwarzschild black hole case (as a reference).

From Fig.\ref{f}, one observes that as the impact parameter \( b \) increases, the deflection of null geodesics grows from weak to strong for all cases, and reaches a peak at \( b = b_{\text{ph}} \) before gradually decreasing thereafter. As \( Q \) increases, \( b_{\text{ph}} \) decreases gradually, while the ranges of 
\( b \) associated with the lensing-ring and photon-sphere regimes expand progressively.

\begin{figure*}[]
\includegraphics[width=1\textwidth]{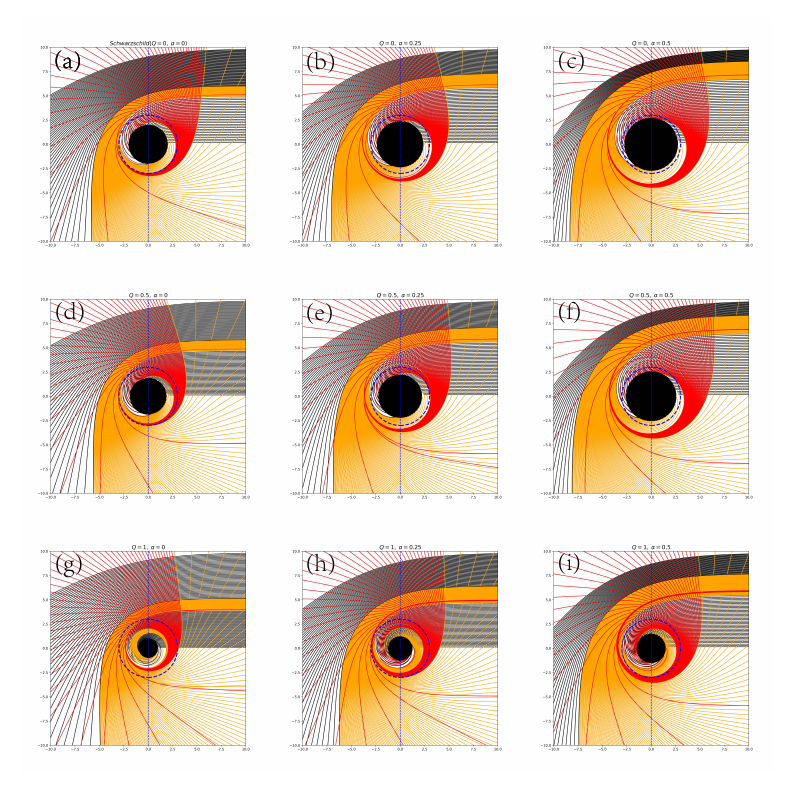}
\caption{
The figure shows photon trajectories around the black hole for electric charges ($Q=0$) (first row), ($0.5$) (second row), and ($1$) (third row), with the Schwarzschild case displayed in the upper-left panel. The $MOG$ parameter ($\alpha$) takes the values ($0$) (first column), ($0.25$) (second column), and ($0.5$) (third column) from left to right. In each panel, the central black disk marks the event horizon, the blue dashed circles indicate the circular photon orbits, and the vertical blue dashed line denotes the equatorial plane on which the accretion disk lies. Photon trajectories are color-coded by the number of intersections with the disk plane: black curves represent direct emission, orange curves correspond to lensing-ring emission, and red curves indicate photon-ring emission.}
\label{g}
\end{figure*}

Subsequently, by applying the null-geodesic equation \eqref{6}, We plot the photon trajectories around the charged $RN$ black hole in $STVG$, as shown in Fig.\ref{g}.

Inspection of Fig.\ref{g} reveals the following key results:

(a)In this figure, the impact parameter \( b \) can be interpreted as the radial offset-measured in the observer’s image plane-between the light ray received at infinity and the black-hole center.

(b)When the charge \( Q \) is fixed,d, the horizon radius \( r_h \), the photon-sphere radius \( r_{\text{ph}} \), and the corresponding critical impact parameter 
\( b_{\text{ph}} \) all increase monotonically as the$ MOG$ parameter $\alpha $ grows. Conversely, for a fixed $\alpha $, these quantities \( r_h \), \( r_{\text{ph}} \) and \( b_{\text{ph}} \) decrease progressively with increasing \( Q \).

(c)As the impact parameter \( b \) increases, the curvature of the photon trajectory first increases and then decreases, peaking at \( b = b_{\text{ph}} \). The trajectory colors follow the sequence black, orange, red, orange, and black.

\begin{figure}[]
\includegraphics[width=0.5\textwidth]{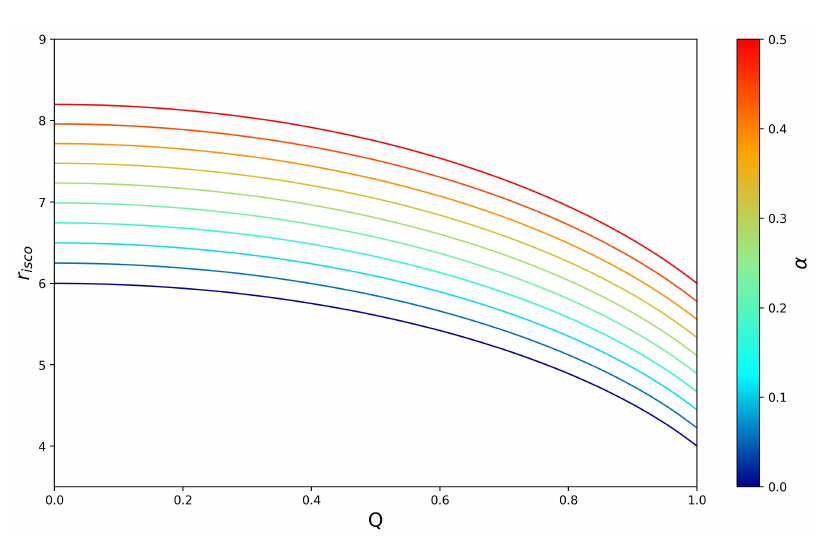}
\caption{
It exhibits the variation law of the innermost stable circular orbit radius \( r_{\text{isco}} \) with the changes in the $MOG$ parameter $\alpha $ and the charge \( Q \).}
\label{h}
\end{figure}

Subsequent calculations allow us to determine the critical values of the impact parameter $b$ associated with different numbers of intersections between photon trajectories and the accretion disk. The results are compiled in Table \ref{table1}, which also lists the event-horizon radius $r_h$, the photon-sphere radius $r_{\text{ph}}$, the critical impact parameter $b_{\text{ph}}$, and the innermost stable circular orbit ($ISCO$) radius $r_{\text{isco}}$. The value of \( r_{\text{isco}} \) can be calculated via Eq.\eqref{14} \cite{Wang:2023vcv}:
\begin{equation}\label{14}
r_{isco} = \frac{ 3f(r_{isco} ) f^{'}( r_{isco} )}{2(f^{'}( r_{isco} ))^{2} - f(r_{isco} ) f^{''}(r_{isco}) } .
\end{equation}
where the prime indicates a derivative with respect to \( r \). Fig.\ref{h} illustrates the variation of the $ISCO$ radius \( r_{\text{isco}} \) with the parameters \( Q \) and $\alpha $.

As can be seen from Fig.\ref{h}, for a fixed value of $\alpha $, \( r_{\text{isco}} \) decreases with an increase in \( Q \), while for a fixed value of \( Q \), \( r_{\text{isco}} \) increases as $\alpha $ rises.

\begin{table*}[]
\centering
\begin{tabular}{p{1.6cm} p{2.7cm} p{1.7cm} p{1.7cm} p{1.7cm} p{1.7cm} p{1.7cm} p{1.7cm} p{1.7cm} p{1.7cm}}
\hline\hline
\rule{0pt}{12pt}$Q$ & $\alpha$ & $b_{2}^{-}$ & $b_{2}^{+}$ & $b_{3}^{-}$ & $b_{3}^{+}$ & $r_{h}$ & $r_{ph}$ & $b_{ph}$ & $r_{isco}$  \\
\hline
\rule{0pt}{12pt}$0$ & $0$ ($Schwarzschild$) & 5.017 & 6.117 & 5.190 & 5.230 & 2.000 & 3.000 & $3\sqrt{3}$ & 6.000 \\
\rule{0pt}{12pt}$0$ & $0.25$ & 5.941 & 7.440 & 6.257 & 6.303 & 2.368 & 3.574 & 6.270 & 7.111 \\
\rule{0pt}{12pt}$0$ & $0.5$ & 6.820 & 8.771 & 7.305 & 7.373 & 2.725 & 4.120 & 7.328 & 8.199 \\
\rule{0pt}{12pt}$0.5$ & $0$ & 4.802 & 5.918 & 4.960 & 4.994 & 1.866 & 2.805 & 4.968 & 5.607 \\
\rule{0pt}{12pt}$0.5$ & $0.25$ & 5.704 & 7.238 & 6.009 & 6.065 & 2.218 & 3.375 & 6.021 & 6.687 \\
\rule{0pt}{12pt}$0.5$ & $0.5$ & 6.561 & 8.557 & 7.035 & 7.125 & 2.561 & 3.921 & 7.064 & 7.750 \\
\rule{0pt}{12pt}$1$ & $0$ & 3.979 & 5.231 & 4.001 & 4.080 & 1.000 & 1.986 & 4.000 & 4.000 \\
\rule{0pt}{12pt}$1$ & $0.25$ & 4.779 & 6.538 & 4.982 & 5.106 & 1.250 & 2.482 & 5.000 & 5.000 \\
\rule{0pt}{12pt}$1$ & $0.5$ & 5.535 & 7.835 & 5.941 & 6.121 & 1.500 & 2.978 & 6.001 & 6.000 \\
\hline\hline
\end{tabular}
\caption{
For the cases where the charge parameter \( Q \) takes values excluding $0$, $0.5$ and $1$, and for $\alpha=0$, $0.25$ and $0.5$, we present the calculated boundary values corresponding to the impact parameter \( b \), along with the results for the black hole event horizon radius $r_{h}$, the photon sphere radius $r_{ph}$, the critical impact parameter $b_{ph}$ and the $ISCO$ radius $r_{isco}$. Note that the scenario with $Q=0$ and $\alpha = 0$ reduce to the Schwarzschild case.
}
\label{table1}
\end{table*}

\subsection{Theoretical constraints on the $MOG$ parameter $\alpha$ and charge \( Q \) based on $ETH$ observational data}\label{4.2}

Using the observed black hole shadow and photon ring radius, we can place theoretical bounds on the $MOG$ parameter $\alpha$ and the charge \( Q \) in an $RN$ black hole metric within the framework of the $STVG$ theory. The $EHT$ Collaboration has investigated the photon ring structures of the two sources $M87^*$ and $SgrA^*$, with the relevant results presented as follows \cite[e.g.][]{EventHorizonTelescope:2019dse,EventHorizonTelescope:2022wok}:

(a)Relevant parameters of the $M87^*$ black hole:

Shadow angular diameter: \( \theta_{\text{M87}^*} = (42 \pm 3) \, \mu\text{as} \),

Distance from Earth: \( D_{\text{M87}^*} = 16.8_{-0.7}^{+0.8} \, \text{Mpc} \),

Mass: \( M_{\text{M87}^*} = (6.5 \pm 0.9) \times 10^9 \, M_\odot \).

(b)Relevant parameters of the Sagittarius $SgrA^*$ black hole:

Shadow angular diameter: \( \theta_{\text{SgrA}^*} = (48.7 \pm 7) \, \mu\text{as} \),

Distance from Earth: \( D_{\text{SgrA}^*} = (8277 \pm 33) \, \text{pc} \),

Mass: \( M_{\text{SgrA}^*} = (4.3 \pm 0.013) \times 10^6 \, M_\odot \).

Based on calculations using general relativity, the diameters of their shadow images are \( d_{\text{sh}}^{\text{M87}^*} = (11 \pm 1.5) \) and \( d_{\text{sh}}^{\text{SgrA}^*} = (9.5 \pm 1.4) \) respectively \cite{Luo:2024avl}.

\begin{figure*}[]
\includegraphics[width=1\textwidth]{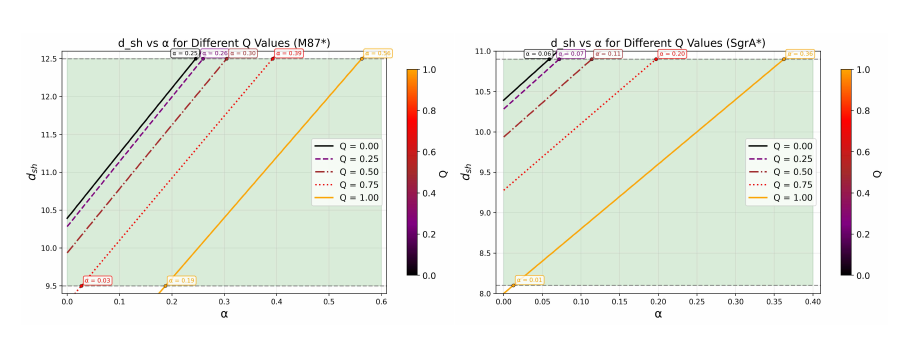}
\caption{
The $EHT$ Collaboration has performed a constraint analysis on the $MOG$ parameter $\alpha$ and charge \( Q \) in the metric adopted in this paper by observing the shadow diameters of the $M87^*$ and $SgrA^*$ black holes. The green shaded region represents the confidence range of the observational values of the black hole shadow diameter, with $M=1$ adopted herein.}
\label{i}
\end{figure*}

As seen by an observer at infinity, the shadow radius $r_{\text{sh}}$ is the lensing-projected image of the circular photon orbit. For a static observer situated at \( r_0 \), the shadow radius can be written as \cite{EventHorizonTelescope:2020qrl}
\begin{equation}\label{15}
r_{\text{sh}} = \frac{r_{\text{ph}}}{\sqrt{-g_{tt}(r_{\text{ph}})}}.
\end{equation}
From Eq.\eqref{1}, we have \( r_{\text{sh}} = r_{\text{ph}} / \sqrt{f(r_{\text{ph}})} \). Expressed in units of the black hole mass, the shadow diameter inferred from observations satisfies \cite{Luo:2024avl}
\begin{equation}\label{16}
d_{\text{sh}} = 2r_{\text{sh}} = \frac{D\theta}{M}.
\end{equation}
Here, \( D \) is the distance from the observer at position \( r_0 \) to the black hole, and \( \theta \) is the angular diameter of the shadow. Fig.\ref{i} presents the constraints imposed by the $EHT$ Collaboration on the $MOG$ parameter $\alpha$ and charge \( Q \) in the metric adopted in this paper, derived from the measured observational data of the shadow diameter.

As can be seen from Fig.\ref{i}, the constraint intervals of the two black holes on the $MOG$ parameter $\alpha$ exhibit a significant discrepancy for different values of \( Q \): for the $M87^*$ black hole, the constraint intervals of $\alpha$ are $[0, 0.25]$, $[0, 0.26]$, $[0, 0.30]$, $[0.03, 0.39]$ and $[0.19, 0.56]$ corresponding to $Q=0$, $0.25$, $0.5$, $0.75$ and $1$, respectively. respectively, whereas for the $SgrA^*$ black hole, the corresponding constraint intervals of $\alpha$ for the same values of \( Q \) are $[0, 0.06]$, $[0, 0.07]$, $[0, 0.11]$, $[0, 0.2]$ and $[0.01, 0.36]$ in turn.A comparison reveals that $SgrA^*$ imposes a stricter upper-limit constraint on the $MOG$ parameter $\alpha$, while $M87^*$ provides a stricter lower-limit restriction on $\alpha$.

Accordingly, the joint constraint interval of the two black holes is adopted in subsequent studies: for $Q=0$, $\alpha \in [0, 0.06]$; for $Q=0.5$, $\alpha \in [0, 0.11]$; and for $Q=1$, $\alpha \in [0.19, 0.36]$. In the case of $\alpha=0$ and $Q=0$, the model recovers the Schwarzschild black hole solution in this limit.

\subsection{Intensity and Optical Appearance of the Reissner-Nordström Black Hole}\label{4.3}

The observational features of black holes arise from the propagation of photons along null geodesics in the strong gravitational field predicted by general relativity. When an accretion disk acts as the radiation source, photons gain energy upon intersecting the accretion disk along null geodesics. The more times such an intersection occurs, the higher the photon energy and the radiation brightness become, leading to an intensity distribution characterized by a bright photon ring, a dark event horizon, and low radiation in the remaining regions of the accretion disk. The photon ring and shadow images of black holes under different conditions are simulated by tuning the parameters $\alpha$ and \( Q \).

Assuming a black hole surrounded by a static, geometrically thin accretion disk whose material radiates isotropically in its rest frame, with emitted frequency $\upsilon_e$, Liouville’s theorem implies that the specific intensity measured by a distant observer is given by \cite[e.g.][]{Gan:2021xdl,Jaroszynski:1997bw,Bromley:1996wb}
\begin{equation}\label{17}
I_o (r, \nu_0) = g^3 I_e (r, \nu_e).
\end{equation}
Here, $g=\nu_0/\nu_e=\sqrt{f(r)}$ denotes the redshift factor, where $\nu_0$ and $\nu_e$ are the observed and emitted frequencies, respectively. The quantities $I_o(r,\nu_0)$ and $I_e(r,\nu_e)$ represent the monochromatic specific intensities of the radiation measured by the observer and emitted by the disk material at radius $r$. Integrating $I_o(r,\nu_0)$ over all frequencies then gives the total observed intensity over the full wavelength range
\begin{equation}\label{18}
\begin{aligned}
I_{\text{obs}} (r) &= \int I_o (r, \nu_0) d\nu_0 \\
&= \int g^4 I_e (r, \nu_e) d\nu_e \\
&= \left(f(r)\right)^2 I_{\text{em}} (r).
\end{aligned}
\end{equation}
Here, \( I_{\text{em}} (r) = \int I_e (r, \nu_e ) \, d\nu_e \) denotes the total emitted intensity of the accretion disk at radius $r$. When the light trajectory tracing back the observer's line of sight intersects the accretion disk, the light extracts energy through interaction with the accretion disk, resulting in an enhancement of the observed brightness. Based on the impact parameter and radiation mechanism, light rays may undergo single or multiple energy acquisitions, leading to an increase in brightness. Due to the geometric properties of geodesics, light rays may intersect the accretion disk multiple times, and the total observed radiation intensity is the superposition of the intensity contributions from each intersection \cite[e.g.][]{Gralla:2019xty,Peng:2020wun}
\begin{equation}\label{19}
I_{\text{obs}} (b) = \sum_{m} \left. \left( f(r) \right)^2 I_{\text{em}} (r) \right|_{r = r_m (b)}.
\end{equation}
By introducing the transfer function \( r_m (b) \), the expression is transformed into a function form with \( b \) as the variable (instead of the original variable \( r \)). This function describes the mapping relationship between the light ray impact parameter \( b \) and the radial coordinate of the \( m \)-th overlap with the accretion disk. The slope \( \frac{dr_m}{db} \) represents the demagnification factor, and the effect becomes stronger as \( m \) increases.

\begin{figure}[]
\includegraphics[width=0.5\textwidth]{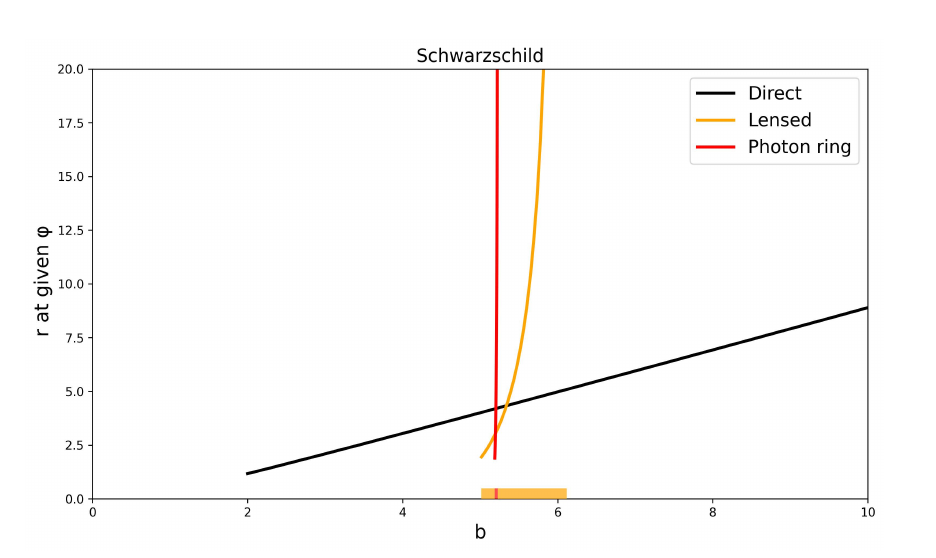}
\caption{
The first three transfer functions for the Schwarzschild solution are represented as follows: the black curve corresponds to the first transfer function, the orange curve to the second, and the red curve to the third.}
\label{j}
\end{figure}

In the intense gravitational field of a black hole, light rays can orbit the black hole multiple times due to the geometric properties of geodesics, passing through the accretion disk repeatedly (e.g., direct emission corresponds to \( m = 1 \), the lensing ring to \( m = 2 \), and the photon ring to \( m \geq 3 \). At each intersection, the light rays extract energy from the accretion disk, contributing to the observed brightness in a linearly superimposed manner, which reflects the cumulative impact of gravitational lensing. Small values of \( m \) (e.g., \( m = 1, 2 \)) correspond to trajectories with weaker lensing effects and contribute to the main brightness (e.g., the lensing ring dominates the total intensity). Large values of \( m \) (e.g., \( m \geq 3 \)) correspond to the photon ring, where the energy extraction fluctuates drastically due to the strong demagnification effect (with a significant slope \( \frac{dr_m}{db} \)). However, their contributions are usually negligible due to their small magnitude \cite{Gralla:2019xty}. The first three transfer functions can be expressed as
\begin{equation}\label{19}
r_1 (b) = \frac{1}{u\left( \frac{\pi}{2}, b \right)}, \quad b \in \left( b_1^-, +\infty \right),
\end{equation}
\begin{equation}\label{20}
r_2 (b) = \frac{1}{u\left( \frac{3\pi}{2}, b \right)}, \quad b \in \left( b_2^-, b_2^+ \right),
\end{equation}
\begin{equation}\label{21}
r_1 (b) = \frac{1}{u\left( \frac{5\pi}{2}, b \right)}, \quad b \in \left( b_3^-, b_3^+ \right).
\end{equation}
Here, \( u(\varphi, b) \) represents the solution to equation \eqref{6}. Fig.\ref{j} presents the distributions of the first three transfer functions corresponding to the Schwarzschild black hole solution.

\begin{figure*}[]
\includegraphics[width=1\textwidth]{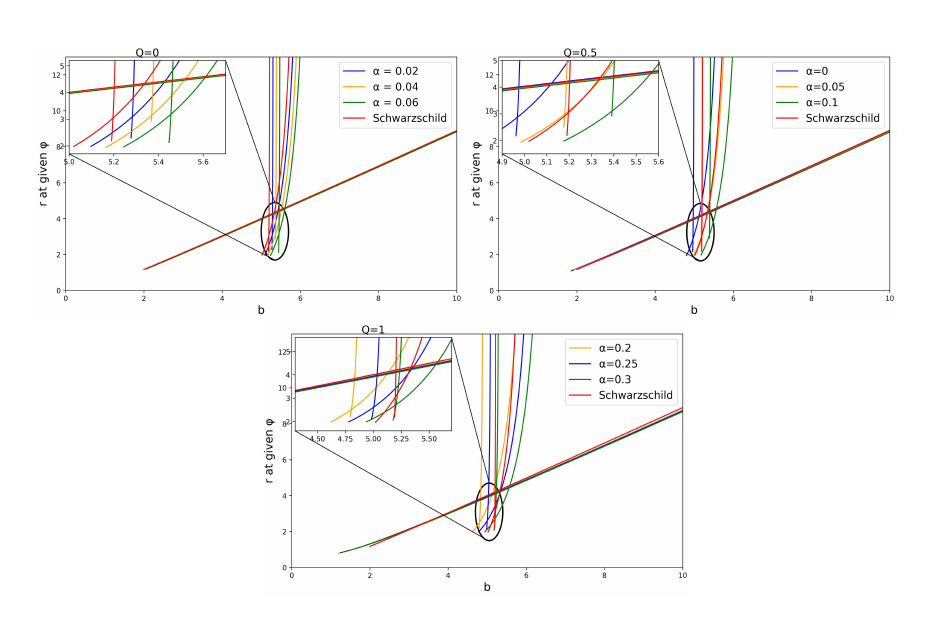}
\caption{
The first three transfer functions for $Q=0$ (with $\alpha=0.02,0.04,0.06$), $Q=0.5$ (with $\alpha=0,0.05,0.1$), $Q=1$ (with $\alpha=0.2,0.25,0.3$), and for the Schwarzschild solution.}
\label{k}
\end{figure*}

As can be seen from Fig.\ref{j}, the contribution sources of the first transfer function include direct radiation, the lensed ring and the photon ring; the second transfer function is contributed only by the lensed ring and the photon ring; and the only contribution source of the third transfer function is the photon ring. Using the parameter constraints obtained in the previous section, Fig.\ref{k} presents the variation laws of the three transfer functions for different values of the parameters $\alpha$ and $Q$.

As can be seen from Fig.\ref{k}, for a fixed value of $Q$, the effective action regions corresponding to the three transfer functions gradually broaden with an increase in $\alpha$. Moreover, the critical impact parameter $b_{ph}$ shifts to the left as $Q$ increases, i.e., for $Q=0$, the case of $\alpha=0.02$ is the closest to the Schwarzschild solution; for $Q=0.5$, $\alpha=0.05$ is the closest; and for $Q=1$, $\alpha=0.3$ is the closest. At the same time, an increase in the intersection number $m$ leads to a significant increase in the slope of the transfer function, which is consistent with the aforementioned conclusion that a larger value of $m$ corresponds to a strong de-magnification effect. In addition, the slopes corresponding to all the first transfer functions $r_{1} (b)$ are approximately equal to $1$, This indicates that the radial coordinate of the first intersection point between the null geodesic and the accretion disk changes approximately linearly with the impact parameter.

To conduct a thorough investigation of the optical appearance characteristics of the $RN$ black hole within the $STVG$ framework, it is essential to define the emission intensity function \( I_{\text{em}} (r) \) of the accretion disk. This function describes how the emission intensity of the accretion disk varies with distance from the black hole. Based on the work of \cite{Li:2021riw}, three specific forms of the emission intensity for toy models are presented as follows:

(a)Quadratic power decay model

The emission intensity decays with radius as a quadratic power, and its expression is
\begin{equation}\label{23}
I_{em}(r)=\begin{cases}I_0\left[\frac{1}{r-(r_{isco}-1)}\right]^2, & r > r_{isco}\\0, & r\leq r_{isco}
\end{cases}, 
\end{equation}

(b)Cubic power decay model

The emission intensity decays at a cubic power rate, with the specific form being
\begin{equation}\label{24}
I_{em}(r)=\begin{cases}I_0\left[\frac{1}{r-(r_{ph}-1)}\right]^3, & r > r_{ph}\\0, & r\leq r_{ph}
\end{cases},
\end{equation}

(c)Slow decay model

The emission intensity decays at a relatively slow rate, modulated by the arctangent function, and its expression is
\begin{equation}\label{25}
I_{em}(r)=\begin{cases}I_0\frac{\frac{\pi}{2}-\tan^{-1}\left[r-(r_{isco}-1)\right]}{\frac{\pi}{2}-\tan^{-1} \left[r_{h}-(r_{isco}-1)\right]}, & r > r_{h}\\0, & r\leq r_{h}\end{cases}. 
\end{equation}
Here, \( r_h \) represents the black hole's event horizon, \( r_{ph} \) is the radius of the photon sphere, and \( r_{isco} \) indicates the radius of the innermost stable circular orbit. The specific values of these parameters are provided in Table \ref{table2}.
\begin{table*}[]
\centering
\begin{tabular}{p{1.8cm}p{1.8cm}p{1.7cm}p{1.7cm}p{1.7cm}p{1.7cm}p{1.7cm}p{1.7cm}p{1.7cm}p{1.7cm}}
\hline\hline
\rule{0pt}{12pt}$Q$ & $\alpha$ & $b_{2}^{-}$ & $b_{2}^{+}$ & $b_{3}^{-}$ & $b_{3}^{+}$ & $r_{h}$ & $r_{ph}$ & $b_{ph}$ & $r_{isco}$  \\
\hline
\rule{0pt}{12pt}$0$ & $0.02$ & 5.097 & 6.214 & 5.277 & 5.311 & 2.030 & 3.028 & 5.283 & 6.090 \\
\rule{0pt}{12pt}$0$ & $0.04$ & 5.165 & 6.326 & 5.354 & 5.390 & 2.060 & 3.077 & 5.369 & 6.180 \\
\rule{0pt}{12pt}$0$ & $0.06$ & 5.244 & 6.428 & 5.437 & 5.480 & 2.090 & 3.127 & 5.455 & 6.269 \\
\rule{0pt}{12pt}$0.5$ & $0$  & 4.764 & 5.920 & 4.952 & 4.995 & 1.866 & 2.805 & 4.968 & 5.607 \\
\rule{0pt}{12pt}$0.5$ & $0.05$ & 4.974 & 6.180 & 5.166 & 5.210 & 1.937 & 2.929 & 5.180 & 5.824 \\
\rule{0pt}{12pt}$0.5$ & $0.10$ & 5.166 & 6.450 & 5.390 & 5.424 & 2.008 & 3.028 & 5.392 & 6.041 \\
\rule{0pt}{12pt}$1$ & $0.20$ & 4.103 & 6.270 & 4.702 & 4.901 & 1.200 & 2.383 & 4.800 & 4.800 \\
\rule{0pt}{12pt}$1$ & $0.25$ & 4.275 & 6.529 & 4.904 & 5.108 & 1.250 & 2.482 & 5.000 & 5.000 \\
\rule{0pt}{12pt}$1$ & $0.30$ & 4.459 & 6.759 & 5.106 & 5.311 & 1.300 & 2.581 & 5.200 & 5.200 \\
\rule{0pt}{12pt}$Schwarzschild$ & & 5.017 & 6.117 & 5.190 & 5.230 & 2.000 & 3.000 & $3\sqrt{3}$ & 6.000 \\
\hline\hline
\end{tabular}
\caption{
The calculated results of the boundary values corresponding to the impact parameter $b$, along with the black-hole event-horizon radius $r_{h}$, photon-sphere radius $r_{ph}$, critical impact parameter $b_{ph}$ and $ISCO$ radius $r_{isco}$, for $Q=0$ (with $\alpha=0.02,0.04,0.06$), $Q=0.5$ (with $\alpha=0,0.05,0.1$), $Q=1$ (with $\alpha=0.20,0.25,0.30$), and the Schwarzschild case.}
\label{table2}
\end{table*}

\begin{figure*}[]
\includegraphics[width=1\textwidth]{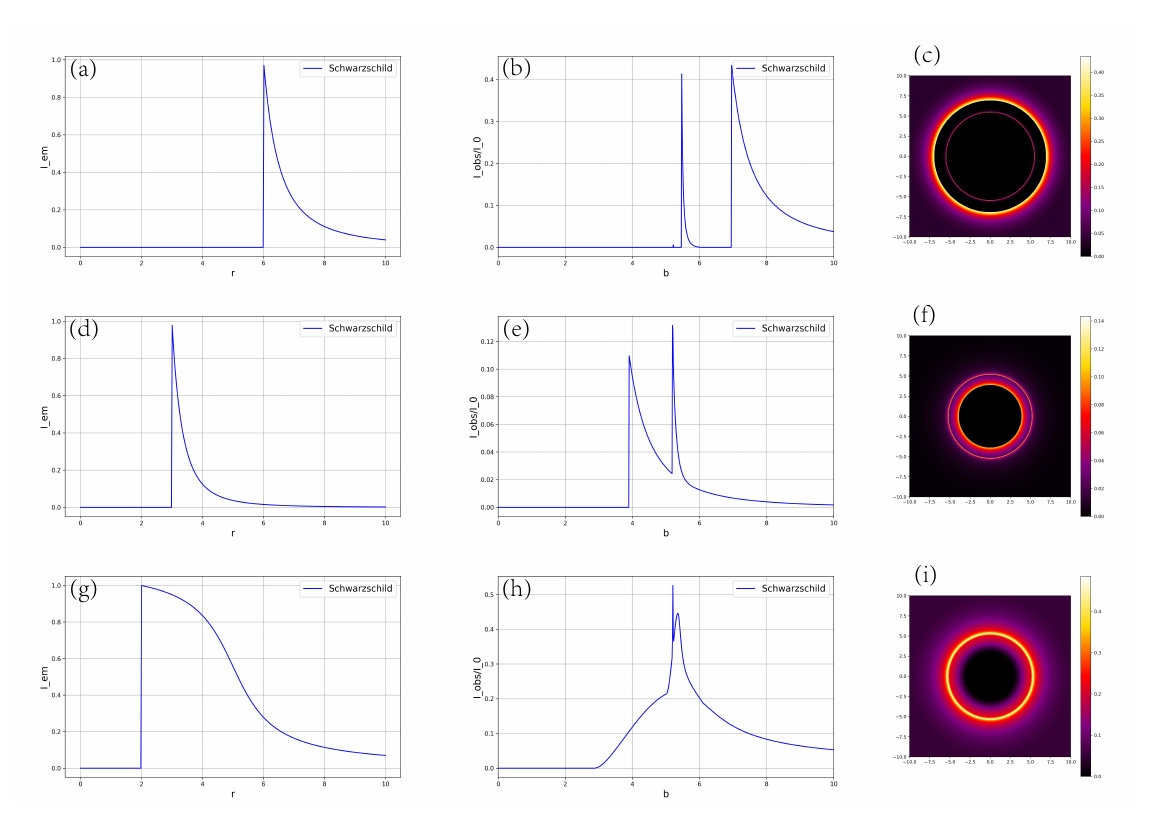}
\caption{
It presents the variation curves of the emission intensity with the radial coordinate \( r \) for the Schwarzschild black hole solution under the three models (left column), the variation curves of the total observed intensity with the impact parameter \( b \) (middle column), along with the optical appearance from the observer's perspective (right column).}
\label{l}
\end{figure*}

By substituting the above three emission intensity functions and the transfer function into Eq.\eqref{19}, for different $MOG$ parameters $\alpha$ and Q, we first plot the variation curves of the emission intensity $I_{em}$ with the radial coordinate \( r \); second, plot the functional curves of the total intensity $\frac{I_{obs}}{I_{0}}$ with the impact parameter \( b \); Then, by exploiting symmetry, we project these curves onto a two-dimensional plane. The optical appearance of the accretion disk is clearly shown through color mapping. Fig.\ref{l} presents the optical appearance of the accretion disk in the vicinity of a Schwarzschild black hole.

As shown in Fig.\ref{l}, among the three models, only the first model displays three characteristic peaks in the total observed intensity. From left to right, the first peak corresponds to the photon ring emission, the second to the lensed ring emission, and the third to the direct emission. Since the intensity of the first peak is very weak and nearly undetectable, its contribution to the total observed intensity can be neglected, with the latter two peaks being dominant.

\begin{figure*}[]
\includegraphics[width=1\textwidth]{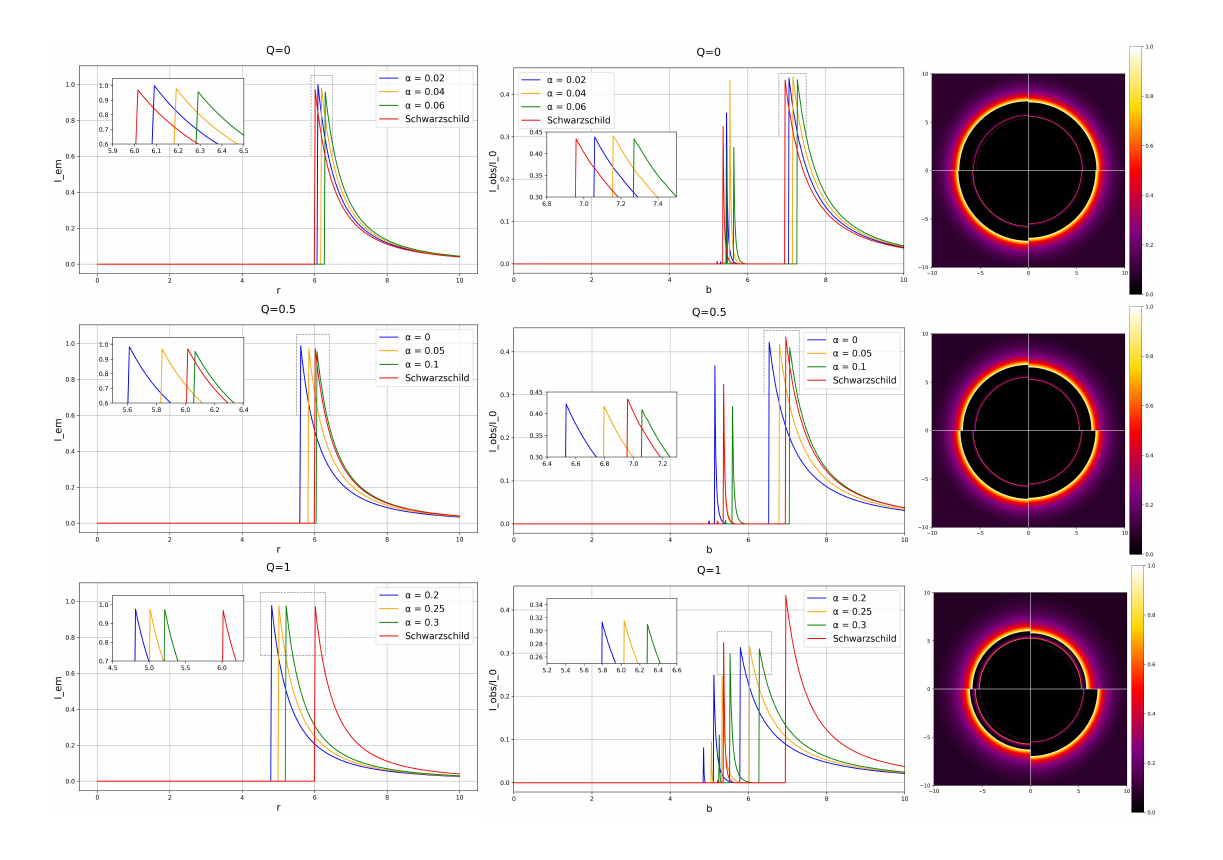}
\caption{
It presents the variation curves of the emission intensity with the radial coordinate \( r \) for the first emission model (left column), the variation curves of the total intensity with the impact parameter \( b \) (middle column), and the optical appearance observed by the observer (right column).}
\label{m}
\end{figure*}

To thoroughly analyze the optical properties of the $RN$ black hole in the context of 
$STVG$ theory, this study systematically investigates the apparent features of the black hole for $Q=0$ (with $\alpha=0.02,0.04,0.06$), $Q=0.5$ (with $\alpha=0,0.05,0.1$), $Q=1$ (with $\alpha=0.2,0.25,0.3$), as well as for the Schwarzschild case. In addition, the optical appearance images are constructed via stitching in our processing, and all color bars are normalized. First, Fig.\ref{m} corresponds to the first model, which visually presents the distribution of the black hole optical appearance for different parameter configurations, thus providing an intuitive basis for the subsequent intensity analysis and exploration of the underlying physical mechanisms.

As shown in Fig.\ref{m}, as $\alpha$ increases, the emission intensity reaches its maximum at the radius of the innermost stable circular orbit for all cases. The total observed intensity exhibits three peaks, with the latter two being dominant: the second peak corresponds to the inner lensed ring, and the third peak to the outer direct ring (located outside the lensed ring). These peaks gradually shift toward larger values of \( b \), with their intensities weakening progressively. The optical appearance reveals a thin lensed ring in the dark region, with a finer photon ring inside it (faintly visible for $Q=0$, $\alpha=0.04$, $Q=1$,$\alpha=0.2$ and $Q=1$,$\alpha=0.3$); a brighter direct ring lies outside the lensed ring, matching the characteristics of the total observed intensity. Additionally, it is observed that a larger value of $\alpha$ leads to a larger photon sphere radius, with no degeneracy observed. This suggests that for the modified $RN$ black hole in the $STVG$ theory, photon rings with different radii correspond to distinct values of $\alpha$ and $Q$. Thus, the properties of this black hole can be studied through the optical appearance of the photon ring.

\begin{figure*}[]
\includegraphics[width=1\textwidth]{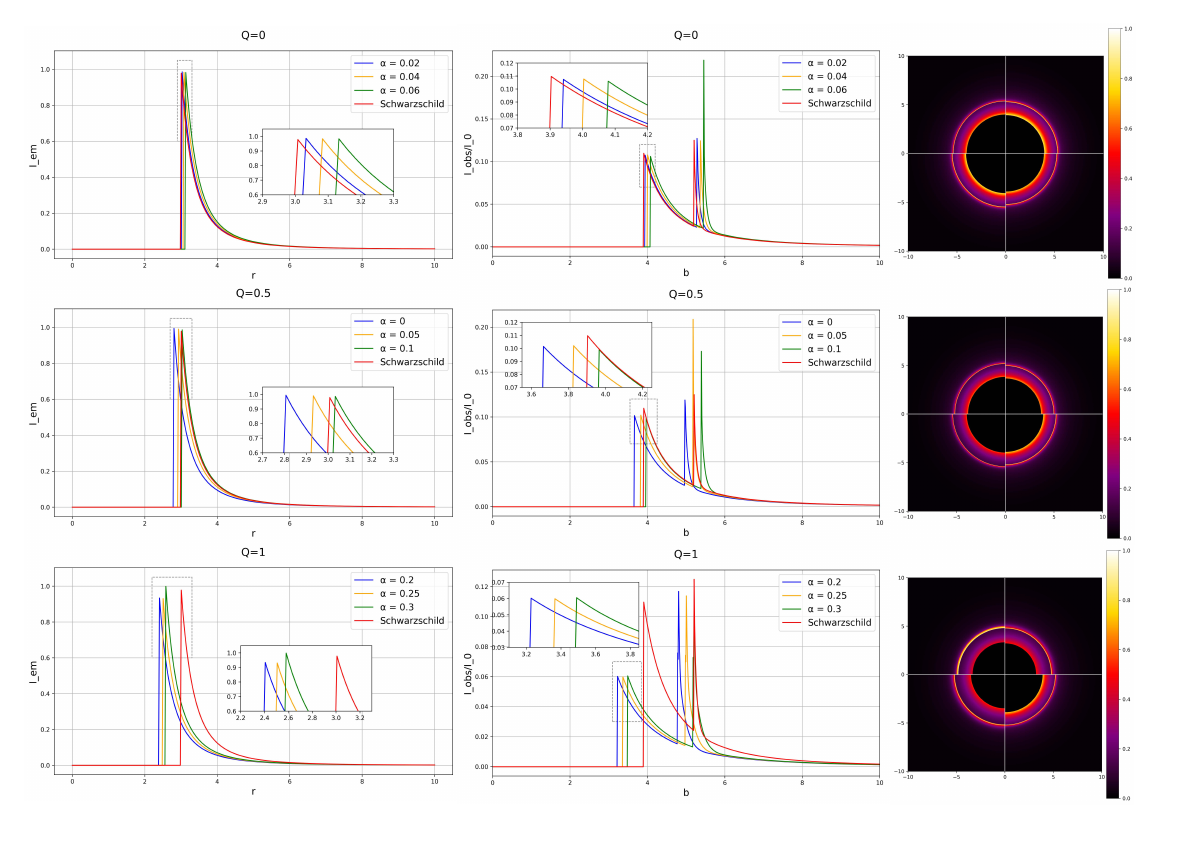}
\caption{
The functional curves of the emission intensity varying with the radial coordinate \( r \) for the second model (left column), the functional curves of the total observed intensity varying with the impact parameter \( b \) (middle column), and the optical appearance as seen by the observer (right column).}
\label{n}
\end{figure*}

We next investigate the variation in the emission intensity with the radial coordinate \( r \), the total observed intensity with the impact parameter \( b \), as well as the black hole optical appearance for the second model, as shown in Fig.\ref{n}.

As can be seen from Fig.\ref{n}, with an increase in $\alpha$, the emission intensity reaches its maximum value at the photon sphere radius for all cases. The total observed intensity exhibits two peaks, with the corresponding optical appearance showing a two-ring structure: the inner ring is the lensed ring and the outer ring is the direct ring. Meanwhile, the peak positions gradually shift toward larger values of \( b \), with their intensities weakening progressively. The optical appearance images show that the inner lensed ring has a higher brightness; in addition, a larger value of $\alpha$ leads to a larger photon sphere radius of the black hole, with no degeneracy phenomenon observed.

\begin{figure*}[]
\includegraphics[width=1\textwidth]{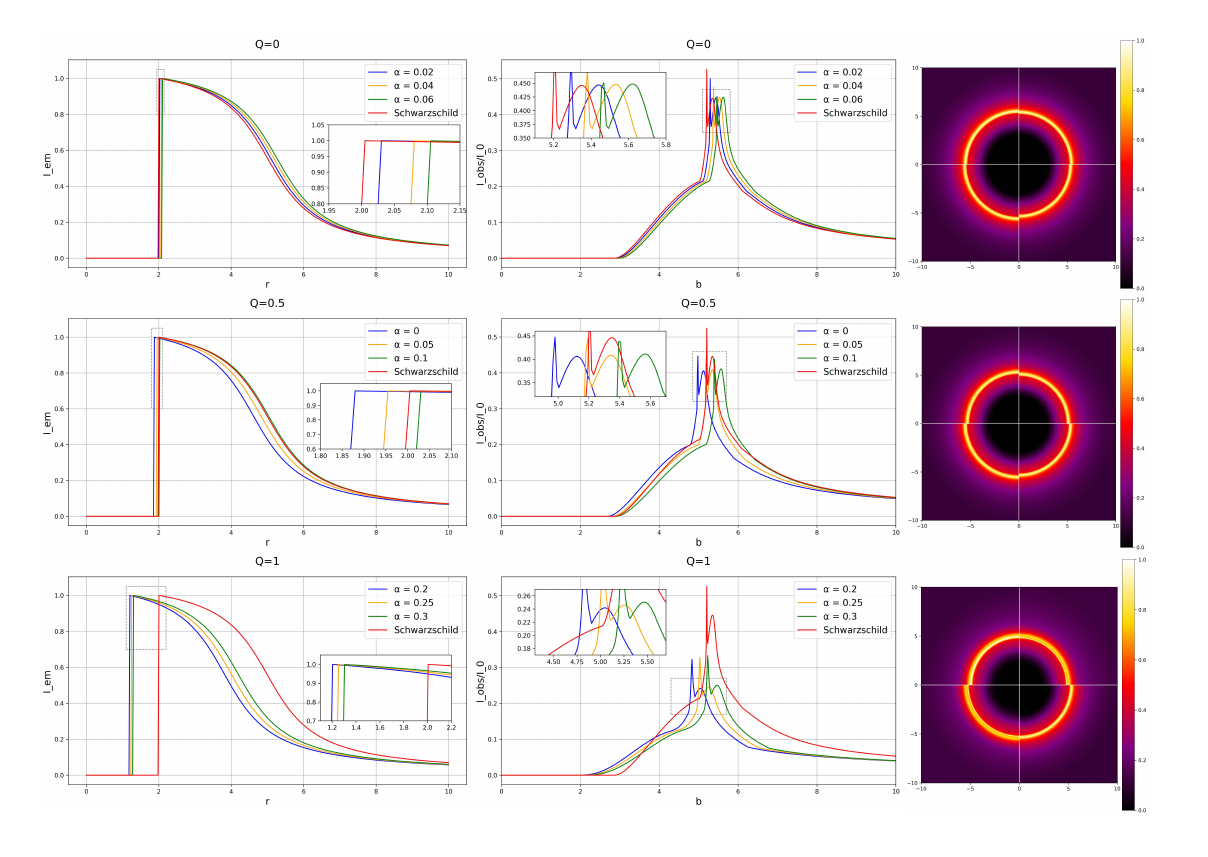}
\caption{
The curves of the emission intensity varying with the radial coordinate \( r \) for the third model (left column), the curves of the total observed intensity varying with the impact parameter \( b \) (middle column), and the optical appearance as seen by the observer (right column).}
\label{o}
\end{figure*}

Fig.\ref{o} presents the variation in the emission intensity with the radial coordinate \( r \), the variation in the total observed intensity with the impact parameter \( b \), as well as the optical appearance of the black hole for the third model.

As shown in Fig.\ref{o}, as $\alpha$ increases, the emission intensity reaches its maximum at the event horizon radius for all cases. The total observed intensity also displays two peaks, but the two-ring structure is hard to distinguish due to the small separation between the peaks, resulting in only a single ring being observed in this model. The peaks gradually shift to larger values of \( b \), with their intensities increasing progressively. The optical appearance images show that within this ring, the brightness varies from dark to bright and back again from the inside out. Additionally, a larger value of $\alpha$ leads to a larger photon sphere radius, with no degeneracy observed.

In summary, the variations in the $MOG$ parameter $\alpha$ and the electric charge $Q$ significantly affect the emission intensity, total observed intensity, and optical appearance of the black hole across the three models. For a fixed $ Q $, the photon sphere radius of the black hole increases as $\alpha$ increases in all cases, with no degeneracy observed. Additionally, the Schwarzschild case can be effectively simulated only for small values of $Q$, offering valuable insights into the study of the $RN$ black hole within the $STVG$ theory.

\section{Conclusions and Discussions}\label{5.0}

This paper primarily examines the spacetime properties of the $RN$ black hole within the $STVG$ theory, derives its geodesic equation and effective potential, and offers a detailed analysis of the fine structure of the photon ring orbit.

In introducing the metric of the $RN$ black hole, we first examine the effects of the $MOG$ parameter $\alpha$ and the electric charge $Q$ on the event horizon. Our results show that as $\alpha$ increases, the event horizon radius also increases, reflecting the regulatory effect of modified gravity on the black hole's horizon position. Additionally, the electric charge $Q$ significantly influences the horizon radius, with curves for different $Q$ values clearly demonstrating the impact of electric charge on the black hole's horizon scale. Based on the cosmic censorship hypothesis (which prohibits the formation of naked singularities), this study selects the $MOG$ parameter range as $[0,0.5]$ and the electric charge $Q$ range as $[0,1]$ to systematically explore the optical appearance properties of the $RN$ black hole within the $STVG$ theory.

Next, by constructing the Lagrangian, we derive the geodesic equations and the effective potential of this black hole, and plot the variation curves of the effective potential with the radial coordinate \( r \) for different values of the $MOG$ parameter $\alpha$ and the electric charge $Q$. Our results show that the event horizon radius of the Schwarzschild black hole is \( r_h = 2 \) and its photon sphere radius is \( r_{ph} = 3 \). Furthermore, it is found that in the region far from the central mass, the effective potential is consistent with the Schwarzschild solution for any values of $\alpha$ and $Q$; in the short-distance range close to the central mass, however, the effective potential deviates from the Schwarzschild solution, and a larger value of $Q$ causes the overall peak to shift to the left (i.e., both the event horizon radius \( r_h \) and the photon sphere radius \( r_{ph}\) decrease gradually).

We use the backward ray-tracing technique to analyze photon motion near the modified black hole under this metric and its interaction with a thin accretion disk in the equatorial plane. First, based on the null geodesic equations, we plot the variation of the intersection number $ n $ between the null geodesics and the accretion disk with the impact parameter $ b $ for different values of $ \alpha$ and $ Q $, as well as the ray trajectory diagrams. It is observed that as the impact parameter $ b $ increases, the curvature of the null geodesics grows from small to large, peaks at $ b = b_{ph} $, and then decreases. With increasing $ Q $, $ b_{ph} $ decreases, and the intervals of the impact parameter corresponding to the lensed and photon rings expand. From the ray trajectory diagrams, we find that for a fixed $ Q $, the event horizon radius $ r_h $, photon sphere radius $ r_{ph} $, and critical impact parameter $ b_{ph} $ all gradually increase with $ \alpha $, while for a fixed  $ \alpha $, these quantities decrease with an increase in $ Q $. Additionally, the photon trajectory color shifts in the order of black, orange, red , orange , black, reflecting changes in curvature. By calculating the radius of the innermost stable circular orbit $( r_{\text{isco}} )$, we observe that for a fixed $ \alpha $, $ r_{\text{isco}} $ decreases with $ Q $, and for a fixed $ Q $, $ r_{\text{isco}} $ increases with $ \alpha $. Theoretical constraints based on observational data from the Event Horizon Telescope $EHT$ show that \( SgrA^* \) imposes a stricter upper limit on $ \alpha $, while \( M87^* \) provides a lower limit. Therefore, the joint constraint interval is used for further analysis: for for $Q=$0, $\alpha \in [0,0.06]$; for $Q=0.5$, $\alpha \in [0,0.11]$; for $Q=1$, $\alpha \in [0.19,0.36]$. Specifically, the model reduces to the Schwarzschild black hole case when $\alpha=0$ and $Q=0$. Finally, by adjusting the parameters $\alpha$ and $ Q $, we simulate the photon ring and shadow images of the $ RN $ black hole in the $ STVG $ theory during the outward radiation process for three toy models. Our results show that, for a fixed $ Q $, as $\alpha$ increases, the positions of the emission intensity peaks in the three models shift, corresponding to the innermost stable circular orbit, photon sphere radius, and event horizon radius. The total observed intensity in the three models presents two distinct peaks: the optical appearance images of the first two models show two bright rings, while only a single bright ring appears in the third model. A consistent trend emerges across all models: for a fixed $ Q $, the photon sphere radius of the black hole increases monotonically with $\alpha$, with no degeneracy observed. Furthermore, the modified black hole model can closely approximate the Schwarzschild black hole case only when both $ Q $ and $\alpha$ are small. These results indicate that the photon sphere radius of the $ RN $ black hole in the $ STVG $ theory corresponds directly to the parameters $\alpha$ and $ Q $, and the black hole's intrinsic properties can be studied through the characteristics of its optical appearance.

This study uncovers the unique physical properties of the photon ring structure of the $RN$ black hole in the $STVG$ theory, and provides an important avenue for the observational verification of quantum gravity theories. The monotonic relationship and non-degeneracy between the $MOG$ parameter $\alpha$, the electric charge $Q$ and the photon sphere radius not only provide an observational criterion for distinguishing different quantum gravity theories \cite[e.g.][]{Saydullayev:2025oop,Yasmin:2025drr,Oteev:2025jvf}, but also exhibit observable signatures of quantum gravitational effects in the strong gravitational field regime. Parameter constraint results based on $EHT$ observational data show that the theoretical predictions of this theory are in good agreement with existing observations \cite[e.g.][]{EventHorizonTelescope:2019dse,EventHorizonTelescope:2022wok}, yet the current observational precision still limits the accurate discrimination of different values of $\alpha$ and $Q$. Future higher-resolution observational facilities, including next-generation $EHT$ \cite{Johnson:2023ynn} and multi-band polarimetric observations \cite{Hadar:2022xag}, are expected to enable the precise measurement of the fine structure of photon rings. However, the static, spherically symmetric model and the simplified accretion disk assumption adopted in this study limit the universality of the results to a certain extent. Future work should be extended to Kerr-type black holes with spin effects taken into account, precise magnetohydrodynamic simulations , and plasma effects \cite[e.g.][]{Dihingia:2024tqr,Koide:2009dt,Jimenez-Rosales:2021ytz,Kocherlakota:2023vff,Dhruv:2024igk}. In the context of $STVG$ theory, the detailed study of the photon ring structure around the $RN$ black hole paves the way for new experimental approaches to address the fundamental problem of quantum gravity unification. With the continuous progress in observational technologies and the refinement of theoretical models, this research is expected to make significant contributions to the validation of quantum gravity theories, the understanding of black holes' quantum nature and the exploration of spacetime's microscopic structure.

\section*{Acknowledgements}
This work was supported by Guizhou Provincial Basic Research Program (Natural Science) (Grant No.QianKeHeJiChu[2024]Young166), the National Natural Science Foundation of China (Grant No.12365008), the Guizhou Provincial Basic Research Program (Natural Science) (Grant No.QianKeHeJiChu-ZK[2024] YiBan027 and QianKeHeJiChuMS[2025]680), the Guizhou Provincial Major Scientific and Technological Program XKBF (2025) 010 (Hosted by Professor Xu Ning), the Guizhou Provincial Major Science and Technological Program XKGF (2025) 009 (Hosted by Professor Xiang Guoyong) and Guizhou Provincial Major Scientific and technological Program (Teacher Fan Lu Lu moderated).
\section*{Appendix}
In this appendix, we provide a brief overview of the derivation process for the geodesic equation \eqref{6}.

Considering the line element corresponding to the metric \eqref{1} (in the equatorial plane, with \( \theta = \pi/2 \) and \( \dot{\theta} = 0 \)) as
\begin{equation}\label{26}
ds^2 = -f(r) dt^2 + \frac{1}{f(r)} dr^2 + r^2 d\varphi^2.
\end{equation}
The null geodesics of photons satisfy \( ds^2 = 0 \), and their equations of motion can be described by the Hamiltonian formalism. Define the Hamiltonian \( H \) as the inner product of the canonical momentum and velocity minus the Lagrangian:
\begin{equation}\label{27}
H = p_t \dot{t} + p_r \dot{r} + p_\varphi \dot{\varphi} - \mathcal{L}.
\end{equation}
where the dot denotes the derivative with respect to \( \tau \). From the Lagrangian
\begin{equation}\label{28}
\mathcal{L} = \frac{1}{2} \left[ -f(r) \dot{t}^2 + \frac{1}{f(r)} \dot{r}^2 + r^2 \dot{\varphi}^2 \right].
\end{equation}
By combining with the definition of canonical momentum \( p_i = \frac{\partial L}{\partial \dot{q_i}} \), the corresponding canonical momentum can be obtained as
\begin{equation}\label{29}
\begin{split}
p_t=\frac{\partial \mathcal{L}}{\partial \dot{t}}=-f(r)\dot{t},\\
p_\varphi  =-\frac{\partial \mathcal{L}}{\partial \dot{\varphi  }} =-r^2\dot{\varphi },\\
p_r=-\frac{\partial \mathcal{L}}{\partial \dot{r}}=-\frac{1}{f(r)}\dot{r},\\
p_\theta =-\frac{\partial \mathcal{L}}{\partial \dot{\theta }} =-r^2\dot{\theta }=0. 
\end{split}   
\end{equation}
Then the Hamiltonian becomes
\begin{equation}\label{30}
H = \frac{1}{2} \left( -f(r) \dot{t}^2 + \frac{1}{f(r)} \dot{r}^2 + r^2 \dot{\varphi}^2 \right).
\end{equation}
For null geodesics, \( H = 0 \) (since \( L = 0 \)), that is
\begin{equation}\label{31}
-f(r) \dot{t}^2 + \frac{1}{f(r)} \dot{r}^2 + r^2 \dot{\varphi}^2 = 0.
\end{equation}
Through further integral of motion of the canonical momentum
\begin{equation}\label{32}
\begin{cases} 
\dfrac{dp_t}{d\tau} = \dfrac{\partial \mathcal{L}}{\partial t} = 0, \\[6pt]
\dfrac{dp_\varphi}{d\tau} = -\dfrac{\partial \mathcal{L}}{\partial \varphi} = 0.
\end{cases}
\end{equation}
That is
\begin{equation}\label{33}
p_t = -f(r)\frac{dt}{d\tau} = \text{const} = E.
\end{equation}
And
\begin{equation}\label{34}
p_\varphi = -r^2 \frac{d\varphi}{d\tau} = \text{const} = L.
\end{equation}
where \( E \) represents the energy of the system, and \( L \) denotes the angular momentum about the axis perpendicular to the invariant plane. From equations \eqref{33} and \eqref{34}, we can derive
\begin{equation}\label{35}
\begin{cases} 
\dot{t} = -\dfrac{E}{f(r)}, \\[6pt]
\dot{\varphi} = -\dfrac{L}{r^2}.
\end{cases}
\end{equation}
Substituting equation \eqref{35} into equation \eqref{31} yields
\begin{equation}\label{36}
\dot{r}^2 = E^2 - \frac{f(r) L^2}{r^2}.
\end{equation}
Treating \( r \) as a function of \( \varphi \) (instead of \( \tau \)), i.e., \( \dot{r} = \frac{dr}{d\varphi} \dot{\varphi} \), and substituting the relevant expressions, we can obtain the equation
\begin{equation}\label{37}
\left( \frac{dr}{d\varphi} \right)^2 = \frac{r^4 E^2}{L^2} - r^2 f(r).
\end{equation}
Let
\begin{equation}\label{38}
r = \frac{1}{u}.
\end{equation}
Simplifying, we get
\begin{equation}\label{39}
\left( \frac{du}{d\phi} \right)^2 = \frac{1}{b^2} - u^2 f\left( \frac{1}{u} \right).
\end{equation}
where \(b = L/E\) is the impact parameter, and the equation above represents the null geodesic equation.


\bibliography{ref}
\bibliographystyle{apsrev4-1}

\end{document}